\documentstyle[epsfig]{elsart}
\newcommand{\lsim}{\mathrel{\rlap{\lower4pt\hbox{\hskip0pt$\sim$}}
\raise1pt\hbox{$<$}}}

\newcommand{\sect}[2]
{\vspace*{0.1\baselineskip}\hspace*{-\parindent}{\bf #1.}~{\bf
#2}\hspace*{\parindent}} 
\newcommand{\sfrac}[2]{\mbox{\footnotesize $\frac{#1}{#2}$}}
\begin{document}
\begin{frontmatter}
{\small Preprint Nos.:~ADP-97-39-T268;~ANL-PHY-8815-TH-97;~MPG-VT-UR~117/97
}

\title{Deconfinement at finite chemical potential}
\author[ua]{A. Bender,}
\author[ua]{G. I. Poulis,}
\author[anl]{C. D. Roberts,}
\author[ur]{S. Schmidt}
\author[ua]{and A. W. Thomas}
\address[ua]{Department of Physics and Mathematical Physics and \\
Special Research Centre for the Subatomic Structure of Matter, \\
University of Adelaide, South Australia 5005, Australia}
\address[anl]{Physics Division, Bldg. 203, Argonne National Laboratory,\\
Argonne IL 60439-4843, USA}
\address[ur]{Fachbereich Physik, Universit\"at Rostock, D-18051 Rostock,
Germany}
\begin{abstract}
In a confining, renormalisable, Dyson-Schwinger equation model of two-flavour
QCD we explore the chemical-potential dependence of the dressed-quark
propagator, which provides a means of determining the behaviour of the chiral
and deconfinement order parameters, and low-energy pion observables.  We find
coincident, first order deconfinement and chiral symmetry restoration
transitions at $\mu_c = 375\,$MeV.  $f_\pi$ is insensitive to $\mu$ until
$\mu \approx \mu_0 \doteq 0.7\,\mu_c$ when it begins to increase
rapidly. $m_\pi$ is weakly dependent on $\mu$, decreasing slowly with $\mu$
and reaching a minimum 6\% less than its $\mu=0$ value at $\mu=\mu_0$.  In a
two-flavour free-quark gas at $\mu=\mu_c$ the baryon number density would be
approximately $3\,\rho_0$, where $\rho_0=0.16\,{\rm fm}^{-3}$; while in such
a gas at $\mu_0$ the density is $\rho_0$.
\end{abstract}
\begin{keyword}
Field theory at finite chemical potential; Pion properties; Confinement;
Dynamical chiral symmetry breaking; Dyson-Schwinger equations\\[2mm]
\end{keyword}
\end{frontmatter}
%
\hspace*{-\parindent}{\bf 1.~Introduction.}\hspace*{\parindent} The
production of a quark-gluon plasma and the quantitative exploration of its
properties is a primary goal of experimental studies of high energy heavy ion
collisions.  Present efforts explore the domain of nonzero baryon number
density~\cite{tdlee95}.  The theoretical study of finite baryon density in
QCD begins with the inclusion of a chemical potential, $\mu$, which modifies
the fermion piece of the Euclidean action: $\gamma\cdot \partial + m \to
\gamma\cdot \partial - \gamma_4 \mu + m$.  Whether conclusions reached via
this application of equilibrium statistical field theory can actually be
explored in heavy ion collisions is an unresolved question.

With the inclusion of $\mu$ 
the fermion determinant acquires an explicit imaginary part, in addition to
those terms associated with axial anomalies.  The $\mu\neq 0$ QCD action
being complex entails that the study of finite density is significantly more
difficult than that of finite temperature, $T$.  As an illustration of this,
whereas numerical simulations of finite-$T$ lattice-QCD have been quite
extensive~\cite{karsch97}, requiring only asymmetric lattices, there is
currently no numerical simulation algorithm for finite density lattice-QCD.
Contemporary studies using the quenched approximation; i.e., eliminating the
fermion determinant, encounter a forbidden region, which begins at $\mu =
m_\pi/2$~\cite{dks96}.  Since $m_\pi\to 0$ in the chiral limit this is a
serious limitation, preventing a reliable study of chiral symmetry
restoration, for example.

At $T=0=\mu$ the Dyson-Schwinger equations [DSEs] have been employed
extensively in the study of confinement and dynamical chiral symmetry
breaking~\cite{dserev}; and the calculation of hadron
observables~\cite{peterrev}.  As the quark-gluon plasma is characterised by
deconfinement and chiral symmetry restoration, they also provide a continuum
tool for the exploration of the onset and properties of this phase of QCD.
The DSEs are a system of coupled integral equations whose solutions, the
$n$-point Schwinger functions, are the fully-dressed Euclidean propagators
and vertices for the theory.  Once all the Schwinger functions are known then
the theory is solved.  To arrive at a tractable problem one must truncate the
system at a given level.  Truncations that preserve the global symmetries of
a field theory are easy to implement~\cite{brs96}.  Preserving the gauge
symmetry is more difficult but progress is being made~\cite{ayse97}.

An often-used approach is to focus on the DSE for the dressed-quark
propagator, $S(p)$: the $2$-point quark Schwinger function, whose kernel is
constructed from the dressed-gluon propagator, $D_{\mu\nu}(k)$, and the
dressed-quark-gluon vertex, $\Gamma_\mu(k,p)$.  Choosing Ans\"atze for
$D_{\mu\nu}(k)$ and $\Gamma_\mu(k,p)$ one obtains a single integral equation
whose solution provides information about quark confinement and dynamical
chiral symmetry breaking (DCSB).  When the Ans\"atze are based on sound
premises, such as the extensive body of DSE studies relating to the gluon
propagator~\cite{dserev,mike} and quark-gluon vertex~\cite{vertex}, and the
results are insensitive to qualitatively equivalent variations of these
Ans\"atze, then the conclusions of such a study can be judged robust.

This approach has been used in a study of deconfinement and chiral symmetry
restoration in 2-flavour QCD at finite-$T$~\cite{bbkr96}.  Therein the quark
DSE was solved using the one-parameter model dressed-gluon propagator of
Ref.~\cite{fr96}, which provided a good description of $\pi$ and $\rho$-meson
observables at $T=0=\mu$, and a continuum order parameter for deconfinement
was introduced.  That study established the existence of coincident,
second-order deconfinement and chiral symmetry restoration transitions at
$T=T_c\approx 150\,$MeV with a critical exponent $\beta=0.33\pm 0.03$, which
is consistent with that of the $N=4$ Heisenberg magnet: $\beta_H=0.38\pm
0.01$.  This has been argued~\cite{krishna} to characterise the universality
class containing 2-flavour QCD.  Both the pion mass, $m_\pi$, and the pion
leptonic decay constant, $f_\pi$, were insensitive to $T$ until $T\approx
0.7\,T_c$.  However, as $T\to T_c$, the pion mass increased substantially, as
thermal fluctuations overwhelmed quark-antiquark attraction in the
pseudoscalar channel, until, at $T=T_c$, $f_\pi\to 0$ and there was no bound
state.  These results confirm those of contemporary numerical simulations of
finite-$T$ lattice-QCD~\cite{karsch97,karsch95}; and make interesting the
exploration of this model at finite chemical potential.

\sect{2}{DSE-model of two-flavour QCD.}  In a Euclidean space formulation,
with $\{\gamma_\mu,\gamma_\nu\}=2 \delta_{\mu\nu}$ and $\gamma_\mu^\dagger =
\gamma_\mu$, the renormalised dressed-quark propagator at $\mu\neq 0$ takes
the form:
\begin{equation}
S(\tilde p )\doteq -i
\vec{\gamma}\cdot \vec{p}\, \sigma_A(\tilde p ) - i \gamma_4\, \omega_p\,
\sigma_C(\tilde p) + \sigma_B(\tilde p)\,, 
\end{equation}
where $\tilde p \doteq (\vec{p},\omega_p\doteq p_4 + i \mu)$,
and satisfies the DSE:
\begin{eqnarray}
S(\tilde p  )^{-1} & \doteq &
i \vec{\gamma}\cdot \vec{p} A(\tilde p  ) +
i \gamma_4 \omega_p C(\tilde p ) + B(\tilde p)\\
& = &
\label{quarkdse}
Z_2^A i \vec{\gamma} \cdot \vec{p} 
+ Z_2 (i\gamma_4 \omega_p + m_{\rm bm})
+ \Sigma^\prime(\tilde p )\,.
\end{eqnarray}
Here $m_{\rm bm}$ is the Lagrangian current-quark bare mass and the
regularised self energy is
\begin{eqnarray}
\Sigma^\prime(\tilde p) & =&
i \vec{\gamma}\cdot \vec{p} \,\Sigma^\prime_A(\tilde p) 
+ i\gamma_4 \omega_p \,\Sigma^\prime_C(\tilde p) 
+ \Sigma^\prime_B(\tilde p)\,; \\
\label{gendse}
\Sigma_{\cal F}^\prime(\tilde p)
&=&  \int^\Lambda_q \sfrac{4}{3} 
g^2 D_{\mu\nu}(\tilde p-\tilde q)\,
\sfrac{1}{4}{\rm tr}\left[{\cal P}_{\cal F} \gamma_\mu S(\tilde q)
\Gamma_\nu(\tilde q,\tilde p)\right]\,,
\end{eqnarray}
where: ${\cal F}= A$, $B$, $C$; ${\cal P}_A \doteq - (Z_1^Ai
\vec{\gamma}\cdot \vec{p}/|\vec{p}|^2) $, ${\cal P}_B \doteq Z_1$, ${\cal
P}_C \doteq - (Z_1i \gamma_4/\omega_p) $; and $\int^\Lambda_q \doteq
\int^\Lambda d^4 q/(2\pi)^4$ represents mnemonically a translationally
invariant regularisation of the integral, with $\Lambda$ the regularisation
mass-scale.  Although not explicitly indicated, the solutions, $\sigma_{\cal
F}$, are functions only of $|\vec{p}|^2$ and $\omega_p^2$.

In renormalising we require that 
\begin{equation}
\left.S(\vec{p},\omega_p)^{-1} 
\right|^{\mu=0}_{|\vec{p}|^2+p_4^2=\zeta^2}
= i\vec{\gamma}\cdot \vec{p} + i\gamma_4 p_4 + m_R(\zeta)\,,
\end{equation}
where $\zeta$ is the renormalisation point and $m_R(\zeta)$ is the
renormalised current-quark mass.  This entails that the renormalisation
constants are:
\begin{eqnarray}
Z_2^A(\zeta^2,\Lambda^2) & = & 1 - \left.\Sigma^\prime_A(\vec{p},p_4)
\right|^{\mu=0}_{|\vec{p}|^2+p_4^2=\zeta^2} \,,\\
Z_2(\zeta^2,\Lambda^2) & = &1 - \left.\Sigma^\prime_C(\vec{p},p_4)
\right|^{\mu=0}_{|\vec{p}|^2+p_4^2=\zeta^2} \,,\\ 
m_R(\zeta^2) & = & Z_2 m_{\rm bm} + \left.\Sigma^\prime_B(\vec{p},p_4)
\right|^{\mu=0}_{|\vec{p}|^2+p_4^2=\zeta^2} \,,
\end{eqnarray}
and yields the renormalised self energies:
\begin{equation}
{\cal F}(\vec{p},\omega_p) = 
\xi_{\cal F} + \Sigma^\prime_{\cal F}(\vec{p},\omega_p) 
        - \left.\Sigma^\prime_{\cal F}(\vec{p},p_4)
                \right|^{\mu=0}_{|\vec{p}|^2+p_4^2=\zeta^2}\,,
\end{equation}
where ${\cal F}=A$, $B$, $C$; $\xi_A = 1 = \xi_C$ and $\xi_B = m_R(\zeta^2)$.

In studying confinement the $\tilde p$-dependence of $A$ and $C$ is
qualitatively important since it can conspire with that of $B$ to eliminate
free-particle poles in the dressed-quark propagator~\cite{brw92}.
Furthermore, in the study of Ref.~\cite{bu1}, the $\tilde p$-dependence of
$A$ and $C$ was a crucial factor in determining the behaviour of bulk
thermodynamic quantities such as the pressure and entropy; being responsible
for these quantities reaching their respective Stefan-Boltzmann limits only
for very large values of $T$ and $\mu$.  It is therefore important in any DSE
study to retain $A(\tilde p)$ and $C(\tilde p)$, and their dependence on
$\tilde p$. 

The DSE-model is specified by the Landau-gauge choice~\cite{fr96}
\begin{eqnarray}
g^2 D_{\mu\nu}(k) & = &
\left(\delta_{\mu\nu} - \frac{k_\mu k_\nu}{k^2}\right)
\frac{{\cal G}(k^2)}{k^2}\,;\\
\label{gksquare}
\frac{{\cal G}(k^2)}{k^2} & = &
4\pi^2 d \left[ 4 \pi^2 m_t^2 \delta^4(k)
+ \frac{1- {\rm e}^{-[k^2/(4 m_t^2)]}}{k^2}\right]\,.
\end{eqnarray}
The first term in (\ref{gksquare}) is an integrable, infrared singularity
that provides long-range effects associated with confinement~\cite{mike}.
The second term ensures that, neglecting logarithmic corrections, the
propagator has the correct perturbative behaviour at large spacelike-$k^2$.
Requiring that the large-distance effects associated with $\delta^4(k)$ are
completely cancelled at small-distances by the second term fixes the ratio of
the coefficients of these two terms in (\ref{gksquare}).  Since ${\cal
G}(k^2)/k^2$ doesn't have a Lehmann representation it can be interpreted as
describing a confined gluon because the nonexistence of a Lehmann
representation is sufficient to ensure the absence of gluon production
thresholds in ${\cal S}$-matrix elements describing colour-singlet to singlet
transitions~\cite{dserev,maris}.

This model, (\ref{gksquare}), has no explicit $\mu$-dependence, which can
arise through quark vacuum polarisation insertions.  As such it may be
inadequate at large values of $\mu$, particularly near any critical chemical
potential.  However, until finite-$\mu$ DSE studies of the gluon propagator
become available no objective assessment of this possibility can be made.  We
therefore advocate proceeding with models such as (\ref{gksquare}) and
assessing the qualitative and quantitative results obtained in the light of
existing experiments and related theoretical studies.

As in Ref.~\cite{fr96} we adopt the additional, simplifying specification:
\begin{equation}
\label{rainbow}
\Gamma_\nu(\tilde q,\tilde p) = \gamma_\nu\,,
\end{equation}
which is often referred to as the ``rainbow approximation''.  Using this
truncation a mutually consistent constraint is: $Z_1=Z_2$ and
$Z_1^A=Z_2^A$~\cite{mr97}.  In $T=0=\mu$ studies this truncation is
quantitatively reliable in Landau gauge; i.e., using (\ref{rainbow}) in
(\ref{gendse}) to obtain $S(\tilde p)$ and calculating bound state masses via
the Bethe-Salpeter equation, the choice of parameters in ${\cal G}(k^2)$ that
gives a good description is little modified by requiring the same quality fit
with a more sophisticated vertex Ansatz, provided it is free of kinematic,
light-cone singularities.  Hence we do not expect the explicit form of the
vertex to have a significant qualitative effect on our conclusions as long as
it is in the class of Ans\"atze prescribed by Ref.~\cite{ayse97}.  This is
supported by the results of Ref.~\cite{brs96}, which indicate that a
systematic improvement of (\ref{rainbow}) has very little effect on
flavour-octet, pseudoscalar meson properties.

In the present context, one can apply the confinement test introduced in
Ref.~\cite{bbkr96} by analysing the configuration-space Schwinger function
\begin{equation}
\Delta_S(\tau) \doteq \frac{1}{2 \pi}\int_{-\infty}^\infty\,dp_4\,
        {\rm e}^{i p_4 \tau}\,\sigma_{B_0}(\vec{p}=0,\omega_p)\,.
\end{equation}
(Here and below the subscript ``0'' denotes a quantity calculated in the
chiral limit, \mbox{$m_R=0$~\cite{mr97}}.)

As an illustration, consider a free, massive fermion, for which
$\sigma_B(\vec{p}=0,\omega_p)= M/(\omega_p^2+M^2)$.  This function has poles
at $p_4^2= -(M\pm \mu)^2$, which are associated with the $\mu$-induced offset
of the particle and antiparticle zero-point energies, and one finds
$\Delta_S(\tau) = 1/2 \exp[-(M-\mu)\, \tau]\, \theta(M-\mu)$.  This function
is positive-definite and monotonically decreasing.\footnote{In the context of
DSE studies the quantity $M$ can be identified with a constituent-quark
mass-scale characteristic of DCSB, which for $u/d$-quarks is $\sim 400\,$MeV.
One therefore expects finite-$\mu$ effects to become noticeable when $\mu\sim
400\,$MeV.}

In contrast, for a Schwinger function with complex-conjugate $p^2$-poles,
$\Delta_S(\tau)$ has zeros at $\tau>0$~\cite{maris}.  Such a Schwinger
function does not have a Lehmann representation and hence can be interpreted
as describing a confined particle, with no associated asymptotic state.  The
confinement order parameter introduced in Ref.~\cite{bbkr96} is derived from
this observation.  Denoting by $\tau_0^z$ the location of the first zero in
$\Delta_S(\tau)$, and defining $\kappa\doteq 1/\tau_0^z$, then deconfinement
is observed if, for some $\mu=\mu_c$, $\kappa(\mu_c) = 0$; at this point the
baryon number density has overwhelmed the confinement mass-scale and the
poles have migrated to the real axis.  Obviously, in the case of a free
particle $\kappa \equiv 0$.

The simplest order parameter for DCSB is
\begin{equation}
\chi = \Re\left[B_0(|\vec{p}|^2=0,\omega_p^2=-\mu^2)\right]\,.
\end{equation}
Its behaviour at any chiral symmetry restoration transition is identical to
that of the chiral-limit vacuum quark condensate.

We solve the DSE for $S(\tilde p)$, (\ref{quarkdse}), numerically with
\begin{equation}
\begin{array}{ccc}
m_t= 0.69\,{\rm GeV}\,,  & & m_R(\zeta) = 1.1\,{\rm MeV}\,,
\end{array}
\end{equation}
which were fixed in Ref.~\cite{fr96} by requiring a best $\chi^2$-fit to a
range of $\pi$-meson observables at $T=0=\mu$.  We use $d=4/9$ in
(\ref{gksquare}) and renormalise at $\zeta = 9.47\,$GeV.  With these choices
our present study has no free parameters.

The pion mass is given by~\cite{fr96}
\begin{eqnarray}
\label{pimass}
m_\pi^2 N_\pi^2 & = &\langle m_R(\zeta) \left(\bar q q\right)_\zeta
\rangle_\pi\,; \\
\langle m_R(\zeta) \left(\bar q q\right)_\zeta \rangle_\pi
& =&
8 N_c \int_p^\Lambda\,B_0\,\left(\sigma_{B_0} 
        - B_0 \left[ \omega_p^2 \sigma_C^2 + |\vec{p}|^2 \sigma_A^2 
                        + \sigma_B^2\right]\right)\,,
\end{eqnarray}
which vanishes linearly with $m_R(\zeta)$, and 
\begin{eqnarray}
\label{npisq}
\lefteqn{N_\pi^2  =  2 N_c \int_p^\Lambda\,
B_0^2\,
\left\{\sigma_A^2 - 2 \left[ 
        \omega_p^2\sigma_C\sigma_C^\prime + 
 |\vec{p}|^2 \sigma_A\sigma_A^\prime + \sigma_B\sigma_B^\prime
        \right] \right.} \\
& & \nonumber \left.\!
   - \sfrac{4}{3}\,|\vec{p}|^2\,\left(
        \left[ \omega_p^2\left(\sigma_C\sigma_C^{\prime\prime} -
        (\sigma_C^\prime)^2\right) + 
        |\vec{p}|^2\left(\sigma_A\sigma_A^{\prime\prime} -
        (\sigma_A^\prime)^2\right) + 
        \sigma_B\sigma_B^{\prime\prime} -
        (\sigma_B^\prime)^2 \right] \right) \right\},
\end{eqnarray}
with $\sigma^\prime_B \equiv
\partial\sigma_B(|\vec{p}|^2,\omega_k^2)/\partial |\vec{p}|^2$, etc. $N_\pi$
is the canonical normalisation constant for the Bethe-Salpeter amplitude in
ladder approximation.  The pion decay constant is obtained from
\begin{eqnarray}
f_\pi\,N_\pi  &= &   4 N_c\int_p^\Lambda\,B_0\,
\left\{ \sigma_A \sigma_B + \sfrac{2}{3} |\vec{p}|^2 
        \left(\sigma_A^\prime \sigma_B - 
         \sigma_A \sigma_B^\prime\right)\right\}\;.
\label{fpi}
\end{eqnarray}

(\ref{pimass})-(\ref{fpi}) were derived~\cite{fr96} under the assumption that
the pion Bethe-Salpeter amplitude is given by $\Gamma_\pi= i \gamma_5 B_0$.
The limitations of this assumption are discussed in Ref.~\cite{mr97} but it
is expected to be qualitatively reliable for small pion masses.  The
difference $\epsilon\doteq |1-f_\pi/N_\pi|$ is a measure of the error
introduced by this representation; and $\epsilon$ is no more than $0.17$ over
the range of $\mu$ considered.  In solving the ladder Bethe-Salpeter equation
with $\Gamma_\pi= i \gamma_5 B_0$, (\ref{pimass}) provides an estimate of the
pion mass obtained thereby that is accurate to within $1$\%. 

\sect{3}{Results and Conclusions.}  It is now clear that in the domain of
confinement and DCSB the solution of the DSE determines the order parameters
for deconfinement and chiral symmetry restoration, and elementary pion
observables such as $m_\pi$ and $f_\pi$.  Since ${\cal F}= {\cal
F}(|\vec{p}|^2,\omega_p^2)$, they are all real.
\begin{figure}[t]
  \centering{\
  \epsfig{figure=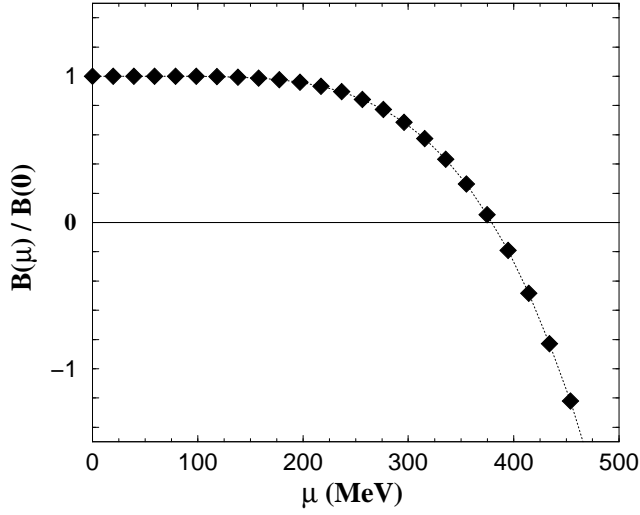,height=7.8cm}}
\caption{${\cal B}(\mu)$ from (\protect\ref{bagpres}); ${\cal B}(\mu)> 0$
marks the domain of confinement and dynamical chiral symmetry breaking.  The
zero of ${\cal B}(\mu)$ is $\mu_c=375\,$MeV.  ${\cal B}(0)= (0.104\,{\rm
GeV})^4$.
\label{bagpressure}}
\end{figure}

To explore the possibility of a phase transition we calculate the difference
between the tree-level auxiliary-field effective-action~\cite{haymaker}
evaluated with the Wigner-Weyl solution, characterised by $B_0\equiv 0$, and
the Nambu-Goldstone solution, characterised by $B_0\not\equiv 0$:
\begin{eqnarray}
\label{bagpres}
\lefteqn{{\cal B}(\mu) = }\\
&& \nonumber 4 N_c \int_p^\Lambda\,
\left\{ \ln\left[\frac{|\vec{p}|^2 A_0^2 + \omega_p^2 C_0^2 + B_0^2}
                {|\vec{p}|^2 \hat A_0^2 + \omega_p^2 \hat C_0^2}\right]
+  |\vec{p}|^2 \left(\sigma_{A_0} - \hat\sigma_{A_0}\right)
+  \omega_p^2 \left(\sigma_{C_0} - \hat\sigma_{C_0}\right)\right\}\,,
\end{eqnarray}
which defines a $\mu$-dependent ``bag constant''~\cite{reg85}.  In
(\ref{bagpres}), $\hat A$ and $\hat C$ represent the solution of
(\ref{quarkdse}) obtained when $B_0\equiv 0$; i.e., when dynamical chiral
symmetry breaking is absent.  This solution exists for all $\mu$.

${\cal B}(\mu)$ is plotted in Fig.~\ref{bagpressure}.  It is positive when
the Nambu-Goldstone phase is dynamically favoured; i.e., has the highest
pressure, and becomes negative when the Wigner pressure becomes larger.  The
critical chemical potential is the zero of ${\cal B}(\mu)$; i.e.,
$\mu_c=375\,$MeV.  This abrupt switch from the Nambu-Goldstone to the
Wigner-Weyl phase signals a first order transition.  The value of $\mu_c$ in
Ref.~\cite{bu1}, obtained without the second term in (\ref{gksquare}); i.e.,
without the ``perturbative tail'', is \mbox{$\approx 30$\%} smaller.

\begin{figure}[t]
 \centering{\
 \epsfig{figure=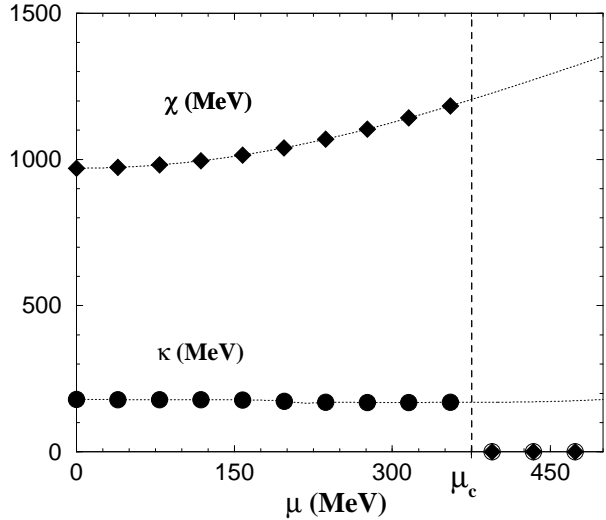,height=7.8cm}}
\caption{The order parameters for chiral symmetry restoration [$\chi$,
diamonds] and deconfinement [$\kappa$, circles].  $\mu_c=375\,$MeV.
\label{figcrit}}
\end{figure}
In Fig.~\ref{figcrit} we plot the order parameters for chiral symmetry
restoration, $\chi(\mu)$, and deconfinement, $\kappa(\mu)$, obtained from our
DSE solutions.  The chiral order parameter {\it increases} with increasing
chemical potential up to $\mu_c $, with $\chi(\mu_c)/\chi(0)\approx 1.2$,
whereas $\kappa(\mu)$ is insensitive to increasing $\mu$.  At $\mu_c$ they
both drop immediately and discontinuously to zero, as expected of a
first-order phase transition.  The increase of the chiral order parameter
with $\mu$ is a necessary consequence of the momentum dependence of the
scalar piece of the quark self energy, $B(\tilde p)$, as is easily seen in
Ref.~\cite{bu1} where the qualitative behaviour of both quantities is
identical.  The vacuum quark condensate behaves in qualitatively the same
manner as $\chi$.


%
\begin{figure}[t]
\centering{\
\epsfig{figure=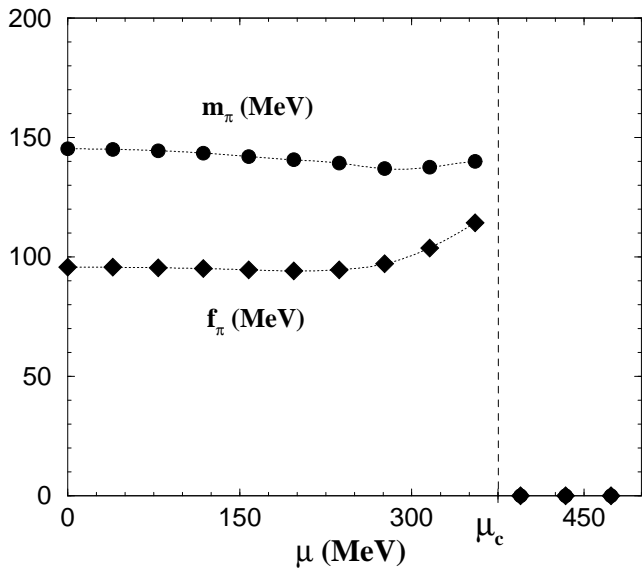,height=8.5cm}}
\vspace*{-\baselineskip}
 
\caption{Chemical potential dependence of the pion mass [$m_\pi$, circles]
and pion leptonic decay constant [$f_\pi$, diamonds].
\label{mpifpi}}
\end{figure}
The behaviour of $m_\pi$ and $f_\pi$ is illustrated in Fig.~\ref{mpifpi}.
One observes that although the chiral order parameter {\it increases} with
$\mu$, $m_\pi$ {\it decreases} slowly as $\mu$ increases.  This slow fall
continues until $\mu \approx 0.7\,\mu_c$, when $m_\pi(\mu)/m_\pi(0) \approx
0.94$.  At this point $m_\pi$ begins to increase although, for $\mu<\mu_c$,
$m_\pi(\mu)$ does not exceed $m_\pi(0)$.  This precludes pion condensation,
in qualitative agreement with Ref.~\cite{kubodera}.  The behaviour of $m_\pi$
results from mutually compensating increases in $\langle m_R(\zeta) (\bar q
q)_\zeta\rangle_\pi$ and $N_\pi^2$.  This is a manifestation of the manner in
which dynamical chiral symmetry breaking protects pseudoscalar meson masses
against rapid changes with $\mu$.  The pion leptonic decay constant is
insensitive to the chemical potential until $\mu\approx 0.7\,\mu_c$, when it
increases sharply so that $f_\pi(\mu_c^-)/f_\pi(\mu=0) \approx 1.25$.  The
relative insensitivity of $m_\pi$ and $f_\pi$ to changes in $\mu$, until very
near $\mu_c$, mirrors the behaviour of these observables at
finite-$T$~\cite{bbkr96}.  For example, it leads only to a $14$\% increase in
the $\pi\to \mu\nu$ decay width at $\mu\approx 0.9\,\mu_c$.  The conjecture
of Ref.~\cite{brown} is inconsistent with the anticorrelation we observe
between the $\mu$-dependence of $f_\pi$ and $m_\pi$.

We expect that improving upon the assumption $\Gamma_\pi= i \gamma_5 B_0$
will only modify these observations to the extent that $f_\pi$ rises slowly
and uniformly on $0<\mu<\mu_c$.  The discussion of pion properties has relied
implicitly upon the ladder Bethe-Salpeter equation.  However,
Ref.~\cite{brs96} indicates that improving upon this truncation; i.e.,
including additional skeleton diagrams in the quark DSE and pion
Bethe-Salpeter equation in a manner that preserves Goldstone's theorem, will
have little effect on pion properties and hence will not qualitatively affect
our conclusions.

The confined-quark vacuum consists of quark-antiquark pairs correlated in a
scalar condensate.  Increasing $\mu$ increases the scalar density: $\langle
\bar q q \rangle$.  However, as long as $\mu<\mu_c$ there is no excess of
particles over antiparticles in the vacuum and hence the baryon number
density remains zero; i.e., $\forall \mu < \mu_c$,
$\rho_B^{u+d}=0$~\cite{bu1}. This is just the statement that quark-antiquark
pairs confined in the condensate do not contribute to the baryon number
density.  After deconfinement the quark pressure increases rapidly, as the
condensate ``breaks-up'', and an excess of quarks over antiquarks develops.
At $\mu\sim 5 \mu_c$ this quark pressure saturates the Stefan-Boltzmann
limit: $P_{u+d}= \mu^4/(2\pi^2)$.

As a {\it gauge} of the magnitude of $\mu_c=375\,$MeV we note that the baryon
number density of a two-flavour free-quark gas at this chemical potential
would be
\begin{equation}
\rho_B^{u_F+d_F}(\mu_c)= \frac{1}{3} \, \frac{2 \mu_c^3}{\pi^2} = 2.9\,\rho_0\,,
\end{equation}
where $\rho_0=0.16\,{\rm fm}^{-3}$ is the equilibrium density of nuclear
matter.  For comparison, the central core density expected in a
$1.4\,M_\odot$ neutron star is $3.6$-$4.1\,\rho_0$~\cite{wiringa}.  Using
this gauge, $\rho_0$ corresponds to $\mu_0\approx 260\,$MeV; i.e., $\mu_0 =
0.7\,\mu_c$.

\vspace*{0.5\baselineskip}\hspace*{-\parindent}{\bf
Acknowledgments.}\hspace*{\parindent} We acknowledge useful conversations
with D. Blaschke, A. H\"oll, P. Maris and A. G. Williams.  For their
hospitality and support during visits where some of this work was conducted:
AB and SS gratefully acknowledge the Physics Division at ANL; CDR, the
Department of Physics at the University of Rostock, the Department of Physics
and Mathematical Physics at the University of Adelaide during a term as a
Distinguished Visiting Scholar and, in a subsequent visit, the Special
Research Centre for the Subatomic Structure of Matter; and GIP, CDR and AWT,
the Institute for Theoretical Nuclear Physics, University of Bonn, during the
``Bonn Workshop on Confinement Physics''.  This work was supported by: the
Australian Research Council; the Department of Energy, Nuclear Physics
Division, under contract no. W-31-109-ENG-38; Deutscher Akademischer
Austauschdienst; the National Science Foundation under grant no. INT-9603385;
and benefited from the resources of the National Energy Research Scientific
Computing Center.


\end{document}